\documentstyle[aps,prl,multicol,psfig]{revtex}

\begin{document}
\draft
\title{Direct Investigation of Superparamagnetism in Co Nanoparticle
Films}

\author{S.I. Woods, J.R. Kirtley, Shouheng Sun, R.H. Koch}
\address{IBM T.J. Watson Research Center; Yorktown Heights, NY 10598}
\date{\today}
\maketitle
\begin{abstract}

A direct probe of superparamagnetism was used to determine the 
complete anisotropy energy distribution of Co nanoparticle films.  
The films were composed of self-assembled lattices of uniform Co 
nanoparticles of 3~nm or 5~nm in diameter, and a variable 
temperature scanning-SQUID microscope was used to measure 
temperature-induced spontaneous magnetic noise in the samples.   
Accurate measurements of anisotropy energy distributions of small volume
samples will be critical to magnetic optimization of nanoparticle devices
and media.

\end{abstract}



\begin{multicols}{2}
\narrowtext


	There has been a recent explosion of research into the 
fabrication and understanding of assemblies of magnetic 
nanoparticles\cite{Sun1,Carpenter,Stamm,Mamiya,Black,Wernsdorfer,Dormann}.
Increasing ability to controllably produce superlattices of these
nanoparticles with uniform material properties will have important
consequences on technology and our basic comprehension of magnetism.
Technologically, nanoparticle films with sharp distributions of magnetic
properties can be used to tailor nanodevices and high-density recording
media with reliable, reproducible characteristics.  Scientifically, a 
uniform array of identical particles is an ideal model system for the 
study of single-particle and interparticle magnetic properties, 
without interference from broad or irregular volume and lattice-spacing
distributions.

    Although most magnetic nanoparticle technological
applications depend on the ferromagnetic state of the particles, 
understanding their superparamagnetic state is crucial to 
controlling their behavior.  For nanoparticles, very small absolute 
deviations in particle size can lead to large changes in blocking 
temperature, interfering with device function at operational 
temperatures.  Fundamental magnetic parameters like anisotropy 
energy and flipping times can be extracted from superparamagnetic 
data\cite{Bean,Petit,Lierop,Jonsson1}, and these parameters can in
turn be used to predict the magnetic response of a sample or device.
Superparamagnetic behavior is a primary enemy to faithful operation
of nanodevices and nanoparticle memories, so it is vital that an accurate
and complete method for extracting magnetic distributions from 
nanoparticle systems be developed.  With such methods, magnetic 
optimization of these technologies will be hastened.

    We have directly probed superparamagnetism in films of
self-assembled cobalt nanoparticles by measuring the spontaneous 
magnetic noise arising from these films as a function of 
temperature, a method some have pursued to measure nanoparticle
properties\cite{Jonsson2}.  Nanoparticle spin flips, induced by thermal
energy, are directly sensed by a micro-SQUID, giving statistical 
information on the magnetic properties of millions of particles in 
the array.  Not only can the average anisotropy energy and width 
be determined, but it is shown below how the entire magnetic anisotropy
energy distribution can be extracted, a complete magnetic fingerprint.  The 
extracted distribution is shown to be self-consistent, closely 
reproducing the data from which it was derived.  The micro-SQUID noise
technique described here is ideally suited for the statistical study of
nanoparticle films because it can determine the anisotropy distribution
without requiring the easy axes of the nanoparticles to be aligned and
has the magnetization sensitivity to investigate submicron area samples
only a nanometer thick.

    The samples investigated were films of Co nanoparticles
deposited on thermally oxidized silicon substrates.  Monodisperse 
Co nanoparticles with the multiply-twinned face-centered cubic 
structure (mt-fcc structure\cite{Ino}) were prepared using
high-temperature solution-phase synthesis\cite{Sun2}.  The size of the
nanoparticles can be controlled by the details of synthesis and
separation techniques, and TEM studies indicate $\sigma~<5\%$
for the diameter distributions of these particles.  Each nanoparticle
was produced with an oleic acid coat of 2~nm
thickness which served to stabilize the particles, prevent oxidation,
and control interparticle spacing.  The nanoparticles were
dispersed in hexane and deposited on oxidized silicon
\begin{figure}
\centerline{\psfig{figure=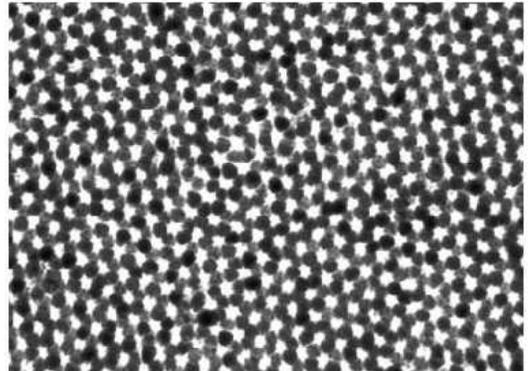,width=2.75in}}
\caption{TEM micrograph of a self-assembled superlattice of 5~nm
diameter Co nanoparticles on silicon-oxide.  An oleic acid coat of 2~nm
around each particle leads to a superlattice spacing of
approximately 9~nm between particle centers.}
\label{fig1}
\end{figure}
\noindent substrates, where the hexane was allowed to slowly
evaporate, leaving 5-10 layers of particles.  The Co nanoparticles
self-assemble on the substrates into close-packed arrays with lattice
spacing approximately (d + 4~nm), where d is the diameter of the 
nanoparticles.  In this study, samples with nominal nanoparticle 
diameters of 3~nm and 5~nm were investigated.  Figure 1 shows a 
TEM image of a superlattice of 5~nm particles on a thin layer
of silicon-oxide.

    Samples were measured using a variable temperature
scanning-SQUID microscope, employing a niobium SQUID with an
integrated 17.8$\mu$m square pickup loop\cite{Kirtley}.
This SQUID was modulated at 100 kHz using a standard flux
modulation and feedback method.  The ac output from the SQUID electronics
was fed into an HP35665A digital signal analyzer to study the
magnetic activity of the nanoparticles at frequencies from 10-10000~Hz
and temperatures from 4-100 K.  The power spectral density of the
SQUID output voltage was recorded and could be easily converted to find the 
magnetic noise power density produced by the sample.  The SQUID could
be moved into and out of contact with the sample, so background signals
and the frequency response of the SQUID could be separated from the 
sample signal.  Measurements were made in a shielded room and 
the sample space had ambient fields of about 1 $\mu$T.  Data were 
taken in near-contact with the nanoparticles ($\sim3~\mu m$ away)
and at heights of 100-500~$\mu$m.
In addition, the magnetization-temperature curves of the samples
were characterized with a Quantum Magnetics MPMS system before
execution of the noise measurement.

    In the simple case of a bistable, single-domain particle with
no interactions, the magnetic noise power measured at a 
temperature T, cyclic frequency $\omega$ and distance d away is
\cite{Kogan,Buckingham}:

\begin{equation}
S_{B}(\omega,T) = \left( \frac{\mu_{o}}{4\pi} \frac{MV}{d^{3}} \right)^{2}
                    \frac{\tau}{1+\omega^{2}\tau^{2}}
\end{equation}
where M is the particle magnetization, V is the particle volume and $\tau$ is 
the average cyclic flipping time for the moment.  This flipping 
time follows the standard activated form $\tau=~\tau_{o}exp(U/k_{B}T)$,
where U is the anisotropy energy of the particle; U=KV, where K is the
effective anisotropy constant.  The noise power is flat at low frequencies 
with a falloff near $\tau^{-1}$.  As a function of temperature, $S_{B}$
displays a single peak centered at $T_{peak}= -U/k_{B}log(\omega\tau_{o})$.
Simply put, $S_{B}(T)$ peaks when the measurement frequency equals the inverse 
flipping time; i.e., when $\omega= \tau^{-1} = \tau_{o}^{-1}exp(-U/k_{B}T)$.

    For a distribution D(U) of particle anisotropy energies, one
finds for the noise power (approximating with use of an average 
prefactor to the Lorentzian in Eq. 1):

\begin{equation}
S_{B}(\omega,T) \propto \int_{0}^{\infty} \frac{\tau_{o}
        e^{U/k_{B}T}}{1+\omega^{2}\tau_{o}^{2}e^{2U/k_{B}T}}
        D(U)dU
\end{equation}
A peaked distribution will exhibit a peaked noise power as a 
function of temperature, just as in the single particle case, but the 
peak will be broadened.  Approximate average values $\bar{\tau_{o}}$ and 
$\bar{U}$ for the flipping time and anisotropy energy can be found by 
assuming this peak occurs at $T_{peak}= -\bar{U}/k_{B}log(\omega
\bar{\tau_{o}})$.

    Results of noise power as a function of frequency from a
film of 5~nm diameter nanoparticles are shown for various 
temperatures in figure 2.  As expected, noise decreases   
\begin{figure}
\centerline{\psfig{figure=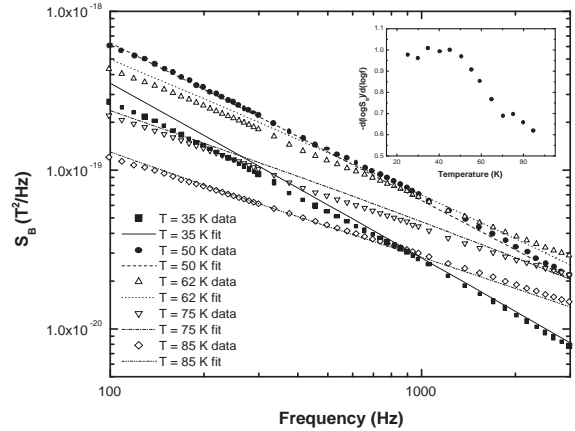,width=3.5in}}
\caption{Magnetic noise power as a function of frequency at
numerous temperatures for a film of 5~nm diameter nanoparticles.  
The fitting curves were generated from an extracted distribution
of anisotropy energies.  The general form of the data is approximately
$S_{B} \propto f^{-\alpha}$, and the inset shows the value of $\alpha$
as a function of temperature for this sample.}
\label{fig2}
\end{figure}
\noindent monotonically in the frequency window for each particular 
temperature, having an approximate form of $S_{B} \propto f^{-\alpha}$.
For the 5~nm nanoparticles, $\alpha$ decreases from about 1.0 to 0.6 as 
temperature is increased from 25 K to 85 K, as shown in the inset 
to figure~2.  Noise power as a function of temperature for both 3~nm
and 5~nm diameter nanoparticle samples is shown in figure~3
for various frequencies.  Each curve peaks at a temperature that 
increases as the measurement frequency increases.  The 3~nm 
sample peaks near T=20~K, and the 5~nm sample peaks near T=55~K.
These temperatures approximate the average
superparamagnetic blocking temperature for each sample and 
agree with the peak location in the zero-field-cooled 
magnetization-temperature curves measured on each sample.
In the following analysis, it is assumed that the noninteracting
form of the equations for $S_{B}$ and $\tau$ (with $\tau_{o}$ and U
possibly renormalized by interactions) is sufficient to describe the
behavior of the system.

    Average values for $\tau_{o}$ and U can be extracted from the
noise data peak locations as discussed above, and a least squares fit 
was made of the log($\omega$) vs. $T_{peak}^{-1}$ data to find these
parameters for the two samples.  For the 3~nm particles, the attempt time 
$2\pi\bar{\tau_{o}}$ and anisotropy energy $\bar{U}$ are $2.16\times10^{-11}$ s
and $5.22\times10^{-21}$ J, respectively.  The average attempt time is
$5.22\times10^{-11}$ s and average anisotropy energy is
$1.35\times10^{-20}$ J for the 5~nm particles.  In the determination
of the full energy distribution to follow, the only input parameter will
be the approximate value calculated for the attempt time\cite{fits}.
\begin{figure}
\centerline{\psfig{figure=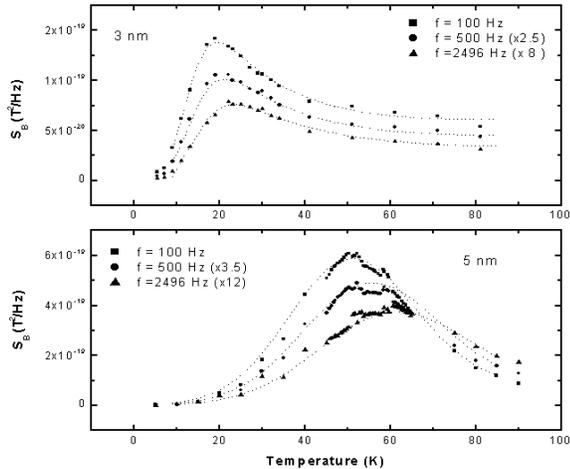,width=3.5in}}
\caption{Magnetic noise power as a function of temperature at
three frequencies for films of 3~nm (top panel) and 5~nm (bottom 
panel) diameter nanoparticles.  The fitting curves were generated 
from extracted distributions of anisotropy energies.}
\label{fig3}
\end{figure}

    By rewriting Eq. 2 as an infinite series, the complete
anisotropy energy can be easily extracted from the noise power 
data under certain conditions.  Let $F(U)=\frac {\tau_{o}e^{U/k_{B}T}}
{1+\omega^{2}\tau_{o}^{2}e^{2U/k_{B}T}}$, then Eq. 2 can be expressed as
$S_{B}(\omega,T) \propto
\int_{0}^{\infty} F(U)D(U)dU$.  F(U) is a function peaked at
$\tilde{U}= -k_{B}Tlog(\omega\tau_{o})$ with width of
approximately $k_{B}T$, falling quickly to zero outside this window.  
D(U) is slowly varying with respect to $k_{B}T$ in this region.  
Expanding D(U) in a power series and evaluating the F(U)-
containing integrals (integrals of the form $\int_{0}^{\infty}
\frac{x^{2n}}{cosh(ax)}dx$ \cite{Gradshteyn}), one finds \cite{Dutta}:

\begin{equation}
S_{B}(\omega,T) \propto \frac{1}{\omega}\sum_{n=0}^{\infty}
        \frac{|E_{2n}|}{(2n)!} \left( \frac {\pi k_{B}T}{2}
        \right)^{2n+1} D^{(2n)} (\tilde{U})
\end{equation}
where $E_{m}$ is the $m^{th}$ Euler number.  As long as D(U) is very 
slowly varying over a scale of $k_{B}T$ around $\tilde{U}$ (the width of 
D(U) is much more than $k_{B}T$ around $\tilde{U}$), the first term alone 
will give a good approximation of the noise power.  Thus, the 
distribution can be found from the noise power according to:

\begin{equation}
D(-k_{B}Tlog(\omega\tau_{o})) \propto \frac {2\omega}{\pi k_{B}T}
        S_{B}(\omega,T)
\end{equation}
If this approximation was not good enough, an iterative procedure
using higher order terms could be used to calculate D(U) from
the noise power.

    The noise power data as a function of temperature for both
the 3~nm and 5~nm samples were used to derive their anisotropy 
energy distributions, according to Eq. 4.  At each frequency the 
distribution was calculated and normalized to make the total 
probability under the distribution equal to unity.  Figure 4 shows 
all the data for numerous frequencies plotted together.  For each 
sample, the data defines a single universal curve, the anisotropy 
energy distribution.  The distribution for the 3~nm particles reaches 
a peak at $4.35\times10^{-21}$~J, has a full-width half-maximum (FWHM) 
of about 110\%, and has a tail at high energies.  For the 5~nm 
particles, the distribution is more symmetric, has a peak at 
$1.22\times10^{-20}$~J, and a FWHM of about 64\%.
\begin{figure}
\centerline{\psfig{figure=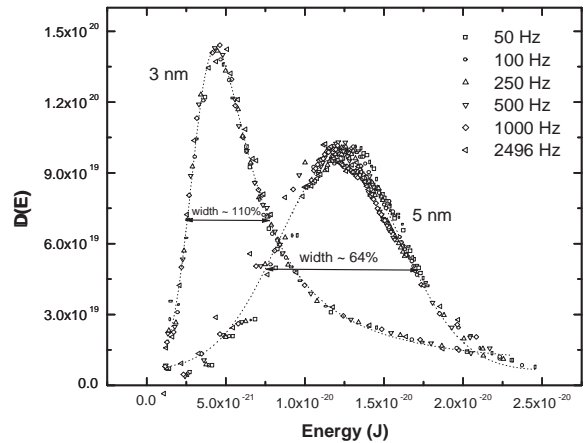,width=3.5in}}
\caption{Extracted anisotropy energy distributions for films of 3~nm
and 5~nm diameter nanoparticles.  The 3~nm curve is peaked at
$4.35\times10^{-21} J$ and exhibits a high energy tail.  The 5~nm distribution 
is peaked at $1.22\times10^{-20} J$ and is symmetric, being well-fit by a 
Gaussian curve (dotted line).}
\label{fig4}
\end{figure}

    To demonstrate the reliability of these extracted
distributions, it is important to show they can reproduce the data 
from which they were derived.  By plugging the derived 
distributions into Eq. 2, one can generate the calculated form for 
the noise power.  The dotted lines in figures 2 and 3 show these 
generated curves, and they provide excellent fits to all the data.

    The derived energy distributions are wider than the apparent
volume distributions, which generally appear Gaussian with $\sigma < 15\%$
(FWHM $< 35\%$).  As it is more difficult to stabilize uniform smaller
particles, it is not surprising that the distribution for the 3~nm
particles is wider than that of the 5~nm particles and shows a tail
indicative of a noticeable fraction of larger diameter particles.  If one 
assumes the energy peak for each sample is that of a
single-domain, spherical nanoparticle of the nominal diameter, then 
$K_{3nm}=3.08\times10^{5} J/m^{3}$ and $K_{5nm}=1.87\times10^{5} J/m^{3}$,
in general agreement with other values derived for mt-fcc Co
nanoparticles\cite{Chen,Jamet}.  It is believed that surface anisotropy
dominates in near-spherical nanoparticles (and is uniaxial, even for
cubic crystal structures), and in this case the anisotropy constant is
expected to scale as the inverse of particle diameter \cite {Bodker1,Bodker2}.
In this study the nominal ratio of the inverse particle diameters of the
two types of samples measured is 1.67, while the ratio in anisotropy
constants is 1.65, in good agreement with the relation expected for
surface anisotropy.  Various reasons could account for energy distributions
significantly wider than nanoparticle volume distributions, including:
existence of nonuniformity in the surface anisotropy; the actively
magnetic regions of the particles do not correspond with their full
TEM-defined sizes; distributions exist in the values of the crystalline,
strain, or shape anisotropies across the nanoparticles; or dipolar
interactions between nanoparticles are significant.

    We investigated the possible effects of dipolar interactions on
the nanoparticle films through Monte Carlo simulations on close-packed
monolayers of particles.  The nanoparticles are given random easy axis
directions and allowed to relax to a low energy state through the
rotation and flipping of nanoparticle magnetizations.  Results indicate
that interparticle dipolar energies are relatively insignificant in the
case of the 3~nm particles but could account for about 15\% of the full
width spread in the flipping energy distribution of the 5~nm particles.
In the case where the 5~nm particles are given a Gaussian diameter
distribution with $\sigma=5\%$, the simulation renders an anisotropy
energy distribution with a width of 51\%, accounting for a significant
part of the 64\% width measured on the 5~nm diameter sample.

    We have presented data gathered by a direct probe of
superparamagnetism in a magnetic nanoparticle system.  The noise 
power technique with a variable temperature micro-SQUID can be used 
to extract precise magnetic energy information on technologically 
important magnetic nanoparticle films and devices of small area and
thickness.  Even for films with sharp volume distributions of nanoparticles,
our magnetic measurement shows that the magnetic distributions can 
be relatively wide.  Thus, such a technique is essential to optimize 
particle growth conditions for magnetic uniformity, critical for any 
application.  A similar scanning-SQUID microscope system employing
a high-$T_{c}$ SQUID could be used to characterize films at room temperature.
The noise measurement technique also holds promise as a sensitive
method to probe and understand interactions in
nanoparticle superlattices as the interparticle spacing is changed.


%
%

\end{multicols}

\begin{references}
\bibitem{fits} In the linear fits, the standard errors for log($\tau_{o}$)
are $\sim11\%$ for the 3~nm particles and $\sim6\%$ for the 5~nm particles,
implying the error for $\tau_{o}$ can be about an order of magnitude.
This possible error only has a small effect on the calculated distribution
results, however.  In the extraction, the input value of $\tau_{o}$ only
affects the energy scaling (multiplies the energy scale by a constant value
at each frequency) and not the shape of the distribution.  A change of 
$\tau_{o}$ by a factor of 10 leads to a change in energy of $\sim13\%$.
\end{references}
\end{document}